\documentclass{article}
\usepackage{amsfonts} \usepackage{amssymb} \usepackage{bbm}
\usepackage{graphicx}  
\usepackage{psfrag}    
\usepackage{amsmath}
\usepackage{amsthm}
\usepackage{enumerate}
\usepackage{diagrams}

\def\idty{{\leavevmode\rm 1\mkern -5.4mu I}} 
\def\norm #1{\Vert #1\Vert}

\mathchardef\ree="023C \mathchardef\imm="023D  

\def\MM{{\mathcal M}}
\def\NN{{\mathcal N}}
\def\BB{{\mathcal B}}
\def\HH{{\mathcal H}}

\def\AA{{\mathcal A}}

\def\veps{{\varepsilon}}
\def\ie{{\it{i.e. }}}

\newtheorem{thm}{Theorem}
\newtheorem{lem}[thm]{Lemma}

\newtheorem{defn}[thm]{Definition}

\newtheorem{examp}[thm]{Examples}

\begin{document}

\title{Tsirelson's Problem}
\author{V. B. Scholz and R. F. Werner\\  \small Institut f\"ur Mathematische Physik\\ \small Technische Universit\"at Braunschweig\\
\small Mendelssohnstr. 3, 38106 Braunschweig, Germany}

\maketitle

\begin{abstract}
The situation of two independent observers conducting measurements on a joint quantum system is usually modelled using a Hilbert space of tensor product form, each factor associated to one observer. Correspondingly, the operators describing the observables are then acting non-trivially only on one tensor factor. However, the same situation can also be modelled by just using one joint Hilbert space, and requiring that all two operators associated to different observers commute, i.e. are jointly measurable without causing disturbance. The problem of Tsirelson is now to decide the question whether all quantum correlation functions between two independent observers derived from by commuting observables can also be expressed using observables defined on a Hilbert space of tensor product form. Tsirelson showed already that the distinction is irrelevant in the case that the ambient Hilbert space is of finite dimension \cite{B.S.Tsirelson:2006fk}. We show here that the problem is equivalent to the question whether all quantum correlation functions can  be approximated by correlation function derived form finite-dimensional systems. We also discuss some physical examples which fulfill this requirement.
\end{abstract}

\newpage

\section{Introduction}

Since the work of John~S.~Bell \cite{Bell1964,Bell1966} we know that quantum mechanics can
violate correlation inequalities, which are valid for arbitrary local
classical theories. On the other hand, quantum mechanics, too,
implies constraints on possible correlations. This was first shown by
Tsirelson \cite{Tsirelson1987}, so just as we can discuss generalized Bell
inequalities, by definition the inequalities satisfied for all
classical local theories, we can investigate generalized {\it
Tsirelson inequalities}, in which the basic assumption is that the
correlations are produced by quantum mechanical systems.
Alternatively, we can look at arbitrary correlation quantities, and
ask for their maximal value according to quantum mechanics, or ask
for the entire convex body of correlations which can be generated by
quantum systems. A still larger convex body is given by all
correlations, satisfying only a basic no-signaling condition. It
also contains the correlations produced by ``non-local boxes'',
hypothetical devices, which nevertheless play an interesting role as
standards of ``non-locality'' and can be used as theoretical devices
in the proofs of other results \cite{Buhrman2003,Fitzi:2008qy,Wolf:2005uq}. These various
correlation sets have received considerable interest in recent years,
as witnessed, for example by Problems 1, 26, 27 on the Braunschweig list \cite{imaphproblist}.
One interesting computer science point of view links the maximal
value of a correlation inequality to the optimal strategy in certain
games with two ``provers'' Alice and Bob, who get some information
(``settings'') from a referee or ``verifier'', and have to respond to
this input without communicating \cite{Ito:2007fj}. Their winnings are determined by
the coincidences in their answers. The various correlation sets then
differ by the resources available to Alice and Bob: For example, if
they can use entangled states, we get the set of quantum
correlations.

In this paper we discuss some possible ambiguities in the definition
of ``correlations produced by quantum systems'', which were noted by
Navascues and Acin \cite{Navascues:2007lq} and formulated in a sharp way by Tsirelson \cite{B.S.Tsirelson:2006fk}. The issue is the notion of ``subsystem'', or the kind of independence
postulated between Alice and Bob. If we just assume that all of
Alice's observables commute with all of Bob's, we might, in
principle, get some larger correlations than if we assume in addition
that these commuting observables act on different tensor factors in a
tensor product decomposition of the underlying Hilbert space.
Tsirelson showed already that if the ambient Hilbert space is finite
dimensional, this distinction is irrelevant. We show here that the approximate version is also true, that is, quantum correlations can be expressed by tensor product subsystems if and only if they can be approximated by correlations between finite dimensional systems.
Hence Tsirelson's problem is the same as the question whether all
quantum correlations can be approximated by correlations between
finite dimensional systems. 

We do not offer a solution, nor even a conjecture. However, we do link the problem to issues well-known in the theory of C*-algebras, von Neumann algebras and operator systems. In this way,
we hope, more methods will become available and, possibly, some
mathematicians specialized on such topics may help to finally
resolve the question.

Our paper is organized as follows: In Sect.~\ref{sec:problem} we
describe the variants of ``quantum correlations'' and state the question in a precise mathematical sense. In Sect.~\ref{sec:opsys} we introduce the notion of an operator system and proof our main result, followed by some physical examples in Sect.~\ref{physimpl}.

\section{Problem statement and results}\label{sec:problem}

The basic scenario of correlation inequalities involves several (in
this paper always two) parties, which can make measurements on parts
of a distributed system. They are free to choose their measuring
device, typically from some finite set of admissible devices. Since
these devices often differ only by the setting of some control
parameter, the devices are also often referred to as {\em settings}.
For a fixed setting it is always clear what possible {\em outcomes}
can be expected from the measurement. Again we assume for simplicity
that the set of outcomes is finite. We denote by $I$ resp. $J$ the finite device sets of Alice and Bob, and to each device we associate a finite outcome set $A_{i}$, $i \in I$ resp. $B_{j}$, $j \in J$. 

Throughout the paper, we are only interested in the case when Alice and Bob want to perform independent measurements, meaning that Bob's choice of a particular measuring device does not depend on Alice's choice. The basic question is now how to model this situation mathematically in the most general way, but assuming quantum mechanics.

Hence, a measurement device is characterized by a set of positive (and hence bounded) operators (POVM) acting on some Hilbert space $\HH$, each associated to a particular outcome, summing up to the identity. Let $\{X\}_{i,\alpha} \subset \BB(\HH)$ denote the collection of positive operators on Alice's side, $i \in I$ identifying the measurement device and $\alpha \in A_{i}$ being the index characterizing a particular outcome. Conversely, let $\{Y\}_{j,\beta} \subset \BB(\HH)$, $j \in J$, $\beta \in B_{j}$ be the set of operators on Bob's side. Here, $\BB(\HH)$ denotes the algebra of all bounded operators on $\HH$. For simplicitiy reasons, we drop the dependence of $\alpha$ and $\beta$ on $i$ and $j$. 

Each measurement device is mapped to a probability distribution by means of a state, \ie a positive  and normalized linear functional $\omega: \BB(\HH) \to \mathbb{R}$ defined on the bounded operators with the additional property that $\omega(\idty) = 1$. In finite dimensions, every state can be descibed by a positive operator $\rho$ having trace one, usually called a density matrix, by means of the identification $\omega(Z) = tr \rho Z$. 

We usually assume that each part of the system can be characterized by a Hilbert space $\HH_{A}$ or $\HH_{B}$. Each measurement operator on Alice's side acts only on $\HH_{A}$, and conversely for Bob. The physical state of the joint system is then described by means of a state $\omega$ on the bounded operators of $\HH_{A} \otimes \HH_{B}$, the Hilbert space tensor product of Alice's and Bob's Hilbert space. The probability $p(i,j | \alpha, \beta)$ that Alice uses measurement device $i$ and detects outcome $\alpha$ while Bob uses measurement device $j$ and detects outcome $\beta$ is described by the expression 
\begin{equation}
p(i,j | \alpha, \beta) = \omega(X_{i, \alpha} \otimes Y_{j,\beta}) \,.
\end{equation}
Although this is the usual model in non-relativistic quantum mechanics, the notion of independent measurements can as well be described by a different setting. The assumption that Alice's choice of a measurement device does not depend on Bob's choice also means that their measurements can both be performed simultaneously, without causing any disturbance. Put differently, this implies that all measurement operators on Alice's side have to commute with Bob's measurement operators. Thus, the same situation can be modelled by using only one Hilbert space $\HH$ and a state $\omega$ on $\BB(\HH)$. Now, Alice and Bob both hold a finite set of positive operators $\{X\}_{i,\alpha} \subset \BB(\HH)$, resp. $\{Y\}_{j,\beta} \subset \BB(\HH)$ with the requirement that $[X_{i,\alpha}, Y_{j,\beta}] = 0$ for all $i,j,\alpha, \beta$. Accordingly, the probability $p(i,j | \alpha, \beta)$ that Alice uses measurement device $i$ and detects outcome $\alpha$ while Bob uses measurement device $j$ and detects outcome $\beta$ is then 
\begin{equation}
p(i,j | \alpha, \beta) = \omega(X_{i, \alpha} \cdot Y_{j,\beta}) \,.
\end{equation}
The obvious question is now wheter both models produce the same set of possible correlation functions. 

Under the assumption that the underlying Hilbert space $\HH$ is finite-dimensional it was shown by Tsirelson that there is no difference between the two models. Without modifications, the proof carries over to the slightly more general case that the sets  $\{X\}_{i,\alpha} \subset \BB(\HH)$ and $\{Y\}_{j,\beta} \subset \BB(\HH)$ generate a finite-dimensional von-Neumann algebra. That is, the set of all linear combinations and products of elements of $\{X\}_{i,\alpha}$ (resp. of $\{Y\}_{j,\beta}$) forms a finite-dimensional vector space. Obvioulsy, this requirement is in particular fulfilled if the underlying Hilbert space $\HH$ is of finite dimension. 

\begin{thm}\label{thmtsi}
Let $\{X\}_{i,\alpha} \subset \BB(\HH)$, and $\{Y\}_{j,\beta} \subset \BB(\HH)$ be finite, commuting sets of positive operators, each generating a finite-dimensional von-Neumann algebra. 

Then there exists a finite-dimensional Hilbert space $\bar{\HH}$ which can be decomposed as $\bar{\HH} = \HH_{A} \otimes \HH_{B}$ such that $\{X\}_{i,\alpha}$ can be mapped isomorphically into $\BB(\HH_{A})$ and correspondingly $\{Y\}_{j,\beta}$ into $\BB(\HH_{B})$.
\end{thm}

\begin{proof}
The proof technique, which we call ``doubling the center'', was already described by Tsirelson \cite{B.S.Tsirelson:2006fk}, but we include it here for the convenience of the reader. Let $\AA_{X}$ resp. $\AA_{Y}$ be the algebra generated by the sets $\{X\}_{i,\alpha}$ resp. $\{Y\}_{j,\beta}$. Obviously, $\AA_{Y}$ lies in the commutant of $\AA_{X}$, $\AA_{X}^{\prime}$. Since $\AA_{X}$ as well as $\AA_{X}^{\prime}$ are finite-dimensional von-Neumann algebras, they can be decomposed into a direct sum of type I factors, $\AA_{X} = \oplus_{k} \AA_{k}$ resp. $\AA_{X}^{\prime} = \oplus_{k} \AA_{k}^{\prime}$. Correspondingly ,the Hilbert space $\HH$ can be decomposed into a direct sum $\HH = \oplus_{k} \HH_{k}$. Now, using the fact that $\AA_{k}$ is of type I for each $k$, we can further decompose each Hilbert space $\HH_{k}$ into a tensor product, $\HH_{k} = \HH^{1}_{k} \otimes \HH^{2}_{k}$ such that $\AA_{k}$ (resp. $\AA_{k}^{\prime}$) acts non-trivially only on $\HH^{1}_{k}$ (resp. $\HH^{2}_{k}$). An elementary proof of this, as well as some physical examples, can also be found in \cite{Zanardi:2004lr}. What remains is to embed the Hilbert space $\HH$ into $\oplus_{k} \HH^{1}_{k} \otimes \oplus_{l} \HH^{2}_{l}$. Thus, $\AA_{X}$ (resp. $\AA_{X}^{\prime}$) can be mapped isomorphically into $\BB(\oplus_{k} \HH^{1}_{k})$ (resp. $\BB(\oplus_{k} \HH^{2}_{k})$).
\end{proof}

Hence, in finite dimension, every quantum correlation function derived from commuting observables can also be represented by observables having tensor product form. In abstract words, this is a consequence of the fact that all von-Neumann algebras represented on a finite-dimensional Hilbert space are of type I. Indeed, as mentioned in the proof, the existence of a tensor decomposition is characteristic for type I von-Neumann algebras. Thus, one might be tempted to conclude that the problem is connected to the possible types of von-Neumann algebras occuring in the description of physical systems. We prove, however, that there exists von-Neumann algebras of any type with the property that the two models described above are still equivalent, meaning that they give rise to the same set of possible correlation functions. 

Summarizing the discussion so far, the question remains open in the case of infinite-dimensional Hilbert spaces. Our main result states that the equivalence of both models still holds if and only if all quantum mechanical correlation functions can be approximated by correlation functions derived from finite-dimensional systems. 

\begin{defn}
Let $p: \bigcup_{i \in I} A_{i} \times \bigcup_{j \in J} B_{j} \to [0,1]$, $(i | \alpha) \times (j | \beta) \mapsto p(i,j | \alpha, \beta)$ be a quantum mechanical correlation function corresponding to two finite sets of measurements $I$, $J$, such that each measurement device $i \in I$ or $j \in J$ posseses a finite set of possible outcomes $A_{i}$ or $B_{j}$. We call $p(i,j | \alpha, \beta)$ \emph{approximately finite dimensional}, if there exist for every $\veps > 0$ a finite dimensional Hilbert space $\HH_{\veps}$, two sets of commuting positive operators $\{X^{\veps}\}_{i,\alpha} \subset \BB(\HH_{\veps})$, $\{Y^{\veps}\}_{j,\beta} \subset \BB(\HH_{\veps})$, summing up to the identity for every fixed $i$ or $j$, and a state $\omega_{\veps}$ on $\BB(\HH_{\veps})$ such that 
\begin{equation}
| p(i,j | \alpha, \beta) - \omega_{\veps}(X^{\veps}_{i,\alpha} \cdot Y^{\veps}_{j,\beta})| < \veps
\end{equation}
holds for all $i,j,\alpha, \beta$.
\end{defn}

The theorem now reads as follows.

\begin{thm}\label{mainthm}
Let $p(i,j | \alpha, \beta) = \omega(X_{i, \alpha} \cdot Y_{j,\beta})$ be a correlation function derived from two finite sets of positive operators acting on a common Hilbert space $\HH$, $\{X\}_{i,\alpha} \subset \BB(\HH)$, $\{Y\}_{j,\beta} \subset \BB(\HH)$ and a state $\omega$ with the additional property that $[X_{i,\alpha}, Y_{j,\beta}] = 0$ for all $i,j,\alpha, \beta$. 

Then the following are equivalent.
\begin{enumerate}[(i)]
\item The correlation function is approximately finite dimensional. 

\item There exist two Hilbert spaces $\displaystyle \HH_{A}$, $\displaystyle \HH_{B}$ and two sets of operators $\displaystyle \{\tilde{X}\}_{i,\alpha} \subset \BB(\HH_{A})$, $\displaystyle \{\tilde{Y}\}_{j,\beta} \subset \BB(\HH_{B})$ and a state $\displaystyle \tilde{\omega}$ such that
\begin{equation}
p(i,j | \alpha, \beta) = \tilde{\omega}(\tilde{X}_{i, \alpha} \otimes \tilde{Y}_{j,\beta})
\end{equation}
holds for all $\displaystyle i,j,\alpha,\beta$.
\end{enumerate}
\end{thm}

As we will see from the proof, most physical models exhibit this approximation requirement. This class includes any fermionic system, quantum spin systems, the CHSH case and usual models from quantum field theory. The discussion of physical models is postponed to section \ref{physimpl}, while the next section is devoted to the proof of the above theorem.  

\section{Operator systems and Tensor norms}\label{sec:opsys}

For starters, we do not restrict ourselves explicitly to the quantum case, but rather study a more general setting. We only assume that there exists a mapping $e$ from the set of outcomes corresponding to all settings to a real linear vector space. More precisely, we introduce for any finite collection of outcome sets $A_{i}$, first the disjoint union $A = \bigcup_{i \in I} A_i$ and then the real vector space $\MM$ spanned by elements $e(\alpha)$ for every outcome $\alpha \in A$. In addition, we assume the existence of a special element $\idty$, such that $\sum_{\alpha \in A_i}e(\alpha)=\idty$. Note that the sum of all outcomes of a given setting is normalized to the same element. 

These requirements reflect the natural assumption that settings and outcomes can be combined to get new devices and that there exists a unit outcome which always gives the answer ``true''. No further dependencies are assumed between the $e(\alpha)$, so that $\dim\MM=1+\sum_i(|A_{i}|-1)$. A general element of $\MM$ can be written as $M=\sum_\alpha m_\alpha e(\alpha)$ and is called positive (``$M\geq0$'') if it has such a representation with $m_\alpha \geq0$. We denote the set of positive elements by $\MM^{+}$. This structure makes $\MM$ an ordered unit vector space. States on $\MM$, which naturally correspond to probability measures on the outcome sets are then defined to be linear, positive and normalized functionals on $\MM$. 

Note that although this construction is somehow similar to the concept of \emph{test spaces} \cite{Barrett:2007qy}, it is different in the sense that the properties ``linearity'' and ``positiveness'' and the concept of an ``identity'' are already included in the definition. However, given any test space, by applying the above procedure we again end up with an ordered unit vector space. 

\subsection{Operator systems}

In quantum theory, each device is given by an observable, i.e. a collection of operators $X_{\alpha}$ labeled by the outcomes $\alpha$ on the system Hilbert space $\HH$ such that $X_{\alpha}\geq0$ and $\sum_\alpha X_{\alpha}=\idty$. Hence, given our measurement space $\MM$, a choice of several quantum observables with the given outcomes is called a {\em quantum representation}, or simply a \emph{representation}, of $\MM$, if it is a linear map $T:\MM\to\BB(\HH)$ such that $M\geq0$ implies $T(M)\geq0$ ($T$ is \emph{positive}) and $T(\idty)=\idty$ ($T$ is \emph{unital}). Clearly, this will lead to observables $X_\alpha=T(e(\alpha))$, when $\alpha \in A_{i}$, and any such choice defines a representation in the sense described. Thus, for each map $T$ we can think of $\MM$ as an linear subspace of $\BB(\HH_{T})$. Such a subspace is also called an \emph{operator system}. More precisely, every self-adjoint subspace of the space of bounded operators on some Hilbert space containing the identity is called an operator system. 

Next, consider two separated labs, in which Alice and Bob each make a
choice of observables. Their basic outcome parameters are again
summarized in an abstract operator system $\MM_A$ for Alice and $\MM_B$ for Bob. They are
operating on separate subsystems, which has the consequence that any
choice Alice might make is compatible with any choice of Bob. Hence,
if Alice chooses an observable with outcomes $A_{i}$ and Bob an
observable with outcomes $B_{j}$, they are measuring together an
observable with outcome set $A_{i} \times B_{j}$. Taking the
disjoint union over $i$ and $j$ we arrive at the total
outcome set $A \times B$, corresponding to the vector space
$\MM_A\otimes\MM_B$, spanned by the elements $e_{A}(\alpha)\otimes e_{B}(\beta)$ and called algebraic tensor product.
Note that the linear normalization relation survives in the tensor
product so we conclude that $\sum_{\alpha \in A_i}e_{A}(\alpha)\otimes
e_{B}(\beta)=\idty\otimes e_{B}(\beta)$ does not depend on the choice of observable
$\alpha$ made by Alice. Hence in this tensor product we have already
encoded the {\it no-signaling condition}. 

Note that the algebraic tensor product $\MM_A\otimes\MM_B$ is not a priori an ordered unit vector space, because we did not define what we mean by positive elements. According to the above discussion, this means that we have not specified how the basis elements $e_{A}(\alpha)\otimes e_{B}(\beta)$ should be represented as observables. Hence, in order to discuss possible \emph{quantum} correlations between Alice and Bob, the choice of the positive cone of $\MM_{A} \otimes \MM_{B}$ turns out to be essential.

\subsection{Tensor norms}

There are two canonical choices for defining positive elements in $\MM_A\otimes\MM_B$. Both are described by defining how the algebraic tensor product is mapped to an operator system, \ie to a space of quantum observables. They are most easily expressed by means of tensor norms. 

\begin{defn}
The p-maximal tensor norm of an element $z \in \MM_{A} \otimes \MM_{B}$, $z = \sum_{k} a_{k} \otimes b_{k}$, $a_{k} \in \MM_{A}$, $b_{k} \in \MM_{B}$ is
\begin{equation}
\norm{z}_{pmax} = \sup\left\{ \norm{T_{A} \cdot T_{B} (z)}_{\BB(\HH)} \, \right\}
\end{equation}
where $T_{A} \cdot T_{B} (z) = \sum_{k} T_{A}(a_{k}) T_{B}(b_{k})$ and the supremum is taken over all positive, unital maps $T_{A} : \MM_{A} \to \BB(\HH)$, $T_{B} : \MM_{B} \to \BB(\HH)$ with commuting ranges. Let $\MM_{A} \otimes_{pmax} \MM_{B}$ denote the normed space obtained from $\MM_{A} \otimes \MM_{B}$ by completion. Accordingly, an element $z \in \MM_{A} \otimes \MM_{B}$ is called positive, if it is mapped to an positive operator for all possible choices of maps $T_{A}$, $T_{B}$ fulfilling the above requirements. That is, if $z = \sum_{k} a_{k} \otimes b_{k}$, then $z > 0$ corresponds to $\sum_{k} T_{A}(a_{k}) \cdot T_{B}(b_{k}) > 0$ for all positive unital maps with commuting ranges.
\end{defn}

This tensor norm catches all cases where the observables are represented as commuting sets of operators. Indeed,the variation principle of Navascues and Acin \cite{Navascues:2007lq} can be reformulated in terms of operator systems using the p-maximal tensor norm. Its  counterpart describes the tensor subsystem situation and is called the p-minimal tensor norm. 

\begin{defn}
The p-minimal tensor norm of an element $z \in \MM_{A} \otimes \MM_{B}$ is
\begin{equation}
\norm{z}_{pmin} = \sup\left\{ \norm{T_{A} \otimes T_{B} (z)}_{\BB(\HH_{A} \otimes \HH_{B})}  \, \right\}
\end{equation}
where the supremum is taken over all positive, unital maps $T_{A} : \MM_{A} \to \BB(\HH_{A})$, $T_{B} : \MM_{B} \to \BB(\HH_{B})$. We define $\MM_{A} \otimes_{pmin} \MM_{B}$ to be the space $\MM_{A} \otimes \MM_{B}$ equipped with this norm. Again, positiveness of an element of the algebraic tensor product $\MM_{A} \otimes \MM_{B}$ is defined according to the mapping properties under any allowed combination of maps $T_{A} \otimes T_{B}$. 
\end{defn}

In both cases we can define the state space as the set of positive linear functionals mapping the identity to one, denoted by $(\MM_{A} \otimes_{pmax} \MM_{B})^{*}$ resp. $(\MM_{A} \otimes_{pmin} \MM_{B})^{*}$. It is evident from the definitions that $\norm{z}_{pmin} \leq \norm{z}_{pmax}$ holds for all $z \in \MM_{A} \otimes \MM_{B}$. 

These two tensor norms precisely catch the possible choices when measuring correlations between subsystems, for example by means of generalized Bell operators. Every such operator is decribed by its decomposition into measuring devices and would thus correspond to an element of the algebraic tensor product $\MM_{A} \otimes \MM_{B}$. The maximal value which can be achieved using quantum states is then exactly equal to its norm. Hence, the choice of the tensor norm characterizes how the corresponding subsystems are modelled, either by just commuting operators or rather by requiring that the underlying Hilbert space is of tensor product form. 

We can also go a step further and define a tensor norm corresponding to the classical setting, \ie the setting where the maps $T_{A}$ and $T_{B}$ are only allowed to take values in commutative algebras. Then we would be able to express the maximal Bell type violation by the difference of two tensor norms. This ``classical'' tensor norm would, however, not respect the fact that $\MM_{A}$ and $\MM_{B}$ are operator systems. More precisely, the normed space obtained from it does not need to be an operator system. Indeed, the p-minimal and p-maximal tensor norms are already the extreme ``quantum'' choices in the sense that every tensor norm in the category of operator systems has to lie in between these two.

Summarizing, in order to show equivalence of the two possible models of subsystem independence, we have to specify the cases where the p-minimal and p-maximal tensor norms are equal. This is the content of our next Lemma, which then directly allows us to prove our main theorem.

\begin{lem}\label{mainlem}
Let $\MM_{A}$ and $\MM_{B}$ be operator systems. The following are equivalent:
\begin{enumerate}[(i)]

\item The p-minimal and p-maximal tensor norm coincide,
\begin{equation}
\MM_{A} \otimes_{pmin} \MM_{B} = \MM_{A} \otimes_{pmax} \MM_{B} \,.
\end{equation}

\item Let $\displaystyle z$ be an element of $\displaystyle \MM_{A} \otimes \MM_{B}$ and $\omega$ a state on $\displaystyle \MM_{A} \otimes_{pmax} \MM_{B}$. There exists for every $\displaystyle \varepsilon >0$ a finite dimensional Hilbert space $\displaystyle \HH_{\varepsilon}$, two representations $\displaystyle T^{\veps}_{A}:\MM_{A} \to \BB(\HH_{\varepsilon})$, $\displaystyle T^{\veps}_{B}:\MM_{B} \to \BB(\HH_{\varepsilon})$ with commuting ranges and a state $\displaystyle \omega_{\varepsilon}$ such that
\begin{equation}\label{eqmainthm}
| \omega(z) -   \omega_{\veps}(T^{\veps}_{A} \cdot T^{\veps}_{B} (z) ) | < \varepsilon \,.
\end{equation}
\end{enumerate}
\end{lem}

\begin{proof}
Assume (ii) and let $x$ be an element of $\MM_{A} \otimes \MM_{B}$ and $\omega$ a state in $(\MM_{A} \otimes \MM_{B})^{*}$. Then there exist Hilbert spaces $\HH_{A}$, $\HH_{B}$ and two completely positive maps $T_{A}:\MM_{A} \to \BB(\HH_{A})$, $T_{B}:\MM_{B} \to \BB(\HH_{B})$ and a state $\omega^{\prime}$ on $\HH_{A} \otimes \HH_{B}$ such that
\begin{equation}
\omega(z) = \omega^{\prime} ( T_{A} \otimes T_{B} (z) ).
\end{equation}
Now let $P_{\varepsilon}$, $Q_{\varepsilon}$ be projections onto finite-dimensional subspaces of $\HH_{A}$, $\HH_{B}$, with the property that $\norm{P_{\varepsilon} T_{A} P_{\varepsilon} \otimes Q_{\varepsilon} T_{B} Q_{\varepsilon} (z) - T_{A} \otimes T_{B} (z)} < \varepsilon$. Of course, the range of $P_{\varepsilon} T_{A}(.) P_{\varepsilon} \otimes \idty$ still commutes with $\idty \otimes Q_{\varepsilon} T_{B}(.) Q_{\varepsilon}$. Furthermore, we have
\begin{equation}
| \omega(z) - \omega ( P_{\varepsilon} T_{A} P_{\varepsilon} \otimes Q_{\varepsilon} T_{B} Q_{\varepsilon} (z) ) | < \varepsilon \,.
\end{equation}
Conversely, suppose that the requirements of (i) are fulfilled. Because any weakly convergent sequence in a finite dimensional Banach space is also convergent in norm, Eq.\eqref{eqmainthm} implies that there exist for every $\veps >0$ two completely positive and unital maps $T^{\veps}_{A}$, $T^{\veps}_{B}$ onto the bounded operators on a finite dimensional Hilbert space with the property that $\norm{T^{\veps}_{A} \cdot T^{\veps}_{B}(z)}_{\BB(\HH_{\veps})} > \norm{z}_{pmax} - \veps$. Now, since the algebras generated by the image of $T^{\veps}_{A}$ resp. $T^{\veps}_{B}$ are acting on a finite-dimensional Hilbert space $\HH_{\varepsilon}$, we can invoke theorem \ref{thmtsi} to find a tensor product decomposition. Hence, we can conclude, 
\begin{align}
\norm{z}_{pmax} - \veps < \norm{T^{\veps}_{A} \cdot T^{\veps}_{B}(z)}_{\BB(\HH_{\veps})} &\leq \norm{(\tilde{T}^{\veps}_{A})\otimes (\tilde{T}^{\veps}_{B})(z)}_{\BB(\HH_{A} \otimes \HH_{B})} \\
 &\leq \norm{z}_{pmin}.
\end{align}
This holds for all $\veps > 0$, which completes the proof.
\end{proof}

\begin{proof}[Proof of Theorem \ref{mainthm}]
Clearly, the choices of observables $\{X\}_{i,\alpha} \subset \BB(\HH)$, $\{Y\}_{j,\beta} \subset \BB(\HH)$ give rise to maps $e_{A}: \bigcup_{i} A_{i} \to \MM_{A}$, $e_{B}: \bigcup_{j} B_{j} \to \MM_{B}$ by means of the identification $\alpha \in A_{i}: e_{A}(\alpha) = X_{i,\alpha}$ and correspondingly for $e_{B}$. The result now follows from Lemma \ref{mainlem} and the fact that the associated state spaces also coincide. 
\end{proof}

\section{Physical Implications and Examples}\label{physimpl}

Usually, physical systems are determined by their observable algebras, being either C*- or von-Neumann algebras. Thus, we have to look for conditions on operator algebras such that all derived correlation functions are approximately finite dimensional. As it can easily be seen, particular examples are nuclear C*-algebras and hyperfinite von-Neumann algebras.

A C*-algebra $\AA$ is called \emph{nuclear} if the following diagram approximately commutes,
\begin{diagram}
          &  & M_{n(\veps)}  &       \\
&\ruTo^{u_{\veps}}  &     &\rdTo^{v_{\veps}}  \\
\AA  &   &\rTo_{id}     &     &\AA
\end{diagram}
that is, if there exists for every $\veps > 0$ two completely positive and unital maps $u_{\veps}$ into some matrix algebra $M_{n(\veps)}$ and $v_{\veps}$ back to $\AA$ which converge pointwise to the identity $id$. Obviously, any measurement space being realized as a subspace of a nuclear C*-algebra will give rise to approximately finite dimensional correlation functions. Thus, commutation relations and tensor form is equivalent in this case. Indeed, another consequence of nuclearity of a C*-algebra $\AA$ is that for any other C*-algebra $\BB$ the tensor algebra $\AA \otimes \BB$ admits only a unique tensor norm, as Choi and Effros have shown \cite{Choi:1978uq,Kirchberg:1977fk}. The class of nuclear C*-algebras contains many physical examples, we just mention a few. 

\begin{examp}
Some nuclear C*-algebras occuring in physics:

\begin{enumerate}

\item All algebras describing fermionic systems, \ie corresponding to the canonical anti-commutation relations.

\item The C*-algebra generated by two projections, corresponding to the case that Alice has two measurement devices with two outcomes each \cite{Raeburn:1989ys}.

\item Uniformly hyperfinite algebras describing infinite spin systems \cite{Keyl:2006ly}.

\end{enumerate}
\end{examp}

For more information about nuclear C*-algebras we recommend the survey by Rordam, \cite{Rordam:2002ve}.

Now, consider the case that  we choose our observables out of a von-Neumann algebra $\NN$. Clearly, the requirements of Theorem \ref{mainthm} are fulfilled if there exists an increasing net of finite dimensional matrix algebras such that the union is weak-*-dense in $\NN$. These von-Neumann algebras, called \emph{hyperfinite}, have been intensively studied in the last decades. There exists many examples of hyperfinite algebras in physics, we name a few.

\begin{examp}
Some hyperfinite von-Neumann algebras occuring in physics:

\begin{enumerate}

\item The unique hyperfinite type $II_{1}$ factor describing the local algebras of an infinite chain of maximally entangled two-qubit states \cite{Keyl:2003wd}.

\item The factors occuring in the construction of algebraic quantum field theory are usually isomorphic to the unique type $III_{1}$ factor \cite{D.-Buchholz:1987kx}.

\end{enumerate}
\end{examp}

Since there exist hyperfinite von-Neumann algebras of any type \cite{Araki:1968lr}, this shows in particular that the problem is not connected to types of von-Neumann algebras. Rather, we have to be able to approximately represent the quantum mechanical correlation functions by finite dimensional systems in order to show the equivalence of commuting observables and those of tensor form.

\section*{Acknowledgments}

V. S. is supported by the European Union through the Integrated Project ``SCALA''.

\bibliographystyle{amsplain}
\bibliography{bib}

\end{document}